\documentclass[letterpaper]{jpconf}
\newcommand{\gcc}{\mathrm{g~cm^{-3} }}
\newcommand{\cms}{\mathrm{cm~s^{-1}}}

\usepackage{graphicx}
\begin{document}
\title{New Approaches for Modeling Type Ia Supernovae}

\author{M. Zingale$^1$, A.~S.~Almgren$^2$, J.~B.~Bell$^2$,
M.~S.~Day$^2$, C.~A.~Rendleman$^2$, and
S.~E.~Woosley$^3$}
\address{$^1$Department of Physics and Astronomy, SUNY Stony Brook, Stony Brook, NY 11794-3800}
\address{$^2$Center for Computational Science and Engineering, Lawrence Berkeley National Laboratory, Berkeley, CA, 94720}
\address{$^3$Department of Astronomy and Astrophysics, University of California, Santa Cruz, Santa Cruz, CA 95064}

\ead{mzingale@mail.astro.sunysb.edu}

\begin{abstract}
Type Ia supernovae (SNe Ia) are the largest thermonuclear explosions
in the Universe.  Their light output can be seen across great
distances and has led to the discovery that the expansion rate of the
Universe is accelerating.  Despite the significance of SNe Ia, there are 
still a large number of uncertainties in current theoretical models.
Computational modeling offers the promise to help answer the
outstanding questions.  However, even with today's supercomputers,
such calculations are extremely challenging because of the wide
range of length and time scales.  In this paper, we discuss several
new algorithms for simulations of SNe Ia and demonstrate some
of their successes.
\end{abstract}

\section{Introduction}

The standard
theoretical picture of a Type Ia supernova is the thermonuclear explosion of
a carbon/oxygen white dwarf that accretes material from a companion
star (see \cite{hillebrandtniemeyer2000} for a good review).  As the
mass of the white dwarf approaches the Chandrasekhar limit---the
maximum mass that can be supported by electron degeneracy
pressure---the temperature and density reach the point where carbon
fusion can take place.  When the energy release due to the burning
exceeds the local cooling rate (due to expansion and neutrino loss),
ignition occurs.  The burning front propagates outward from the center
of the star as a thermonuclear flame, wrinkling via the
Rayleigh-Taylor instability and interactions with turbulence.  As it
continues to burn through the star, the flame accelerates to a large
fraction of the sound speed (and possibly transitions into a
detonation), producing large amounts of nickel.  The energy release is
sufficient to unbind the star.

Large-scale computing has already led to advances in the theoretical
understanding of Type Ia supernovae (SNe Ia); however, much more
is still unknown.  
An accurate model of SNe Ia needs to capture physical processes from
the scale of the carbon flame, $O(10^{-5}~\mathrm{cm}) -
O(10~\mathrm{cm})$, to the scale of the star, $O(10^8~\mathrm{cm})$.
The temporal scales are equally impressive, from the century of
convection leading up to the ignition of the flame to the seconds-long
explosion.  This is truly a multiscale
problem, and even at the dawn of petascale computing, a fully
resolved simulation of all these processes is beyond the available resources.

Simulations of SNe Ia can be broken down roughly into two
types---large eddy simulations operating on the scale of the star
(see
\cite{roepkehillebrandt2005,gamezo2005,plewa:2004,garciasenz:2005}),
and small-scale simulations that resolve the thermonuclear flame
width and model a small, $ < O(100~\mathrm{cm})$, region of the star
(see \cite{niemeyerhillebrandt1995,roepke2003,SNrt,SNrt3d}).  
These two types of simulations are complementary in constructing 
a full picture of the explosion.

The large-scale simulations follow the evolution of the full star down
to typical resolutions of 1~km.  On scales smaller than this, they
rely on subgrid models to describe the physics of the flame
propagation.  The majority of research groups
performing full star explosion simulations use the fully compressible
piecewise parabolic method \cite{ppm} for the hydrodynamics.
Additionally, they require a way to represent and advance the flame
front---typically either thickened flames (see, e.g.,
\cite{khokhlov:1995}) or level-sets \cite{levelset}.  These
simulations have successfully showed that a thermonuclear
carbon deflagration can release enough energy to unbind the star.
However, these pure deflagration models tend to produce weak
explosions and leave behind too much unburned carbon.  A transition to
detonation at the late stages of the explosion has been proposed
\cite{niemeyerwoosley1997,khokhlov1997}, and may produce more
energetic explosions \cite{gamezo2005}.  Detonations are not without
their problems, however \cite{niemeyer1999}.  In particular, it is
unknown if a detonation can develop at all in these environments.
This is one of the major outstanding problems in the modeling of SNe Ia.

Another major uncertainty concerns the initial conditions for the explosion.  
The spatial distribution of the hot spots that seed the
flame can have an enormous impact on whether the explosion is
successful \cite{niemeyer:1996,plewa:2004,garciasenz:2005}.  In
practice, large-scale calculations begin by randomly
initializing one or more hot spots on the grid and propagating the
flame outward from there.  In reality, the white dwarf is convecting
from the century or more of smoldering burning that precedes the
ignition \cite{Woosley:2004}, and burning fronts can ignite over a finite time interval.
With few exceptions \cite{livne:2005}, the large scale calculations ignore the
pre-existing convective velocity field.
Long-time three-dimensional simulations of the convective period are required 
to generate more realistic initial conditions for the ignition and 
resultant explosion.

Simulations on the scale of the flame can be used to formulate and
calibrate the subgrid models employed by the full-star calculations.
As the flame propagates outward from the center of the star,
it encounters lower density fuel and the laminar flame speed
decreases and the flame becomes thicker.  For most of its life, the
flame is in the flamelet regime---characterized by a sharp interface
between the fuel and the ash.  At densities less than $3\times
10^7~\gcc$, the flame becomes broad enough that turbulent eddies on
this scale can disrupt the structure of the flame directly, without
burning away.  At this point, a mixed region of fuel and ash develops,
and the flame is said to be burning in the distributed burning regime.
Small-scale simulations of the distributed burning regime can
help answer the question of whether a transition to detonation 
is possible late in the evolution of the explosion.

In this paper, we discuss new algorithms that allow for efficient
simulation of both the small scales and the full star.  Our approach
exploits the fact that the pre-explosion evolution and flame
propagation are subsonic;  only in the very late stages of the
explosion does the Mach number approach unity.  Filtering sound waves
allows for much longer time evolution than fully compressible codes
and forms the basis for a new generation of SNe Ia evolution codes.

\section{Small-scale flame modeling}

\subsection{Low Mach number approach}

The laminar flame speed of a thermonuclear carbon flame is very
subsonic, with Mach numbers less than $10^{-2}$.  Efficient
simulation of turbulent flames on small scales requires a code well-suited 
to low Mach number flows.   The algorithm described below was developed
originally for low-speed terrestrial combustion, which shares
the common feature that the flame speed and fluid speed are both much
less than the speed of sound.
On small scales, the background stratification of the star is negligible, and 
the pressure in the domain can be assumed constant.  By expanding the
state variables in powers of Mach number, $M,$ the pressure can be
decomposed into a dynamic component, $\pi$, and thermodynamic
component, $p_0$, the ratio of which is $O(M^2)$.  In the low Mach
number model we replace the total pressure, $p$, by $p_0$
everywhere except in the momentum equation; this substitution
decouples pressure and density variations and has the effect of
removing sound waves from the system.  The thermodynamic constraint
that $p = p_0$ can be recast, by differentiating the equation
of state along particle paths,
as a constraint on the velocity field:
\begin{equation}
\vec{\nabla} \cdot \vec{U} = S
\label{eq:sn_constraint}
\end{equation}
In this equation, the source term, $S$, represents the compressibility effects due
to thermonuclear energy release and thermal diffusion.  


\begin{figure}[t]
\centering
\includegraphics[width=3.25in]{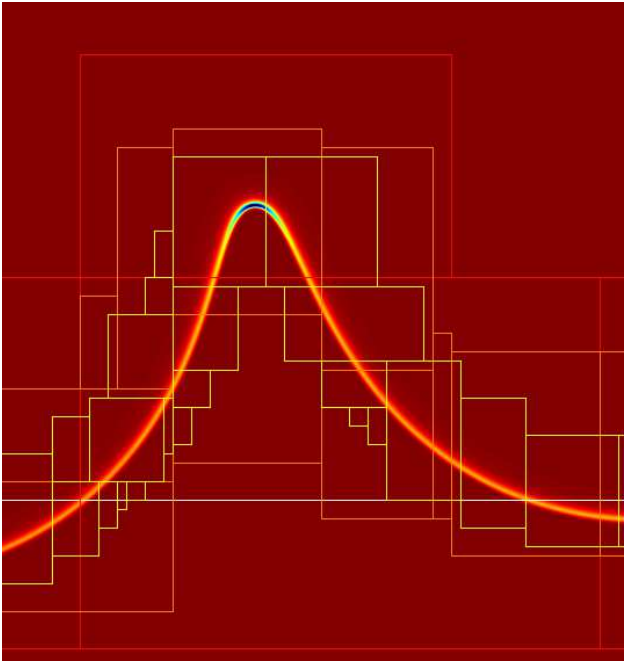} \hspace{0.1in}
\begin{minipage}[b]{2.5in}
\caption{\label{fig:amr} Close-up of a Rayleigh-Taylor unstable flame
showing the nuclear energy generation rate.  The adaptive mesh
refinement algorithm efficiently refines around the flame.\\}
\end{minipage}
\end{figure}

We solve the low Mach number system using a second-order
accurate approximate projection method originally developed for 
solving the incompressible Navier-Stokes equations \cite{almgren-iamr}, 
extended to low Mach number combustion \cite{DayBell:2000}, and explained in
detail for SNe Ia in \cite{Bell:2004}.  Since sound waves are filtered from the
system, the timestep is limited only by the bulk velocity, not the
soundspeed.  For low Mach number flows, this means that a factor of $\sim 1/M$
fewer timesteps are needed relative to fully compressible formulations.  

The low Mach number formulation is solved in an 
adaptive mesh refinement framework in order to focus
the spatial resolution on the regions of the flow that are of
particular interest, such as a flame front.
The gains in efficiency due to the
combination of the low Mach number formulation and adaptive mesh refinement
have opened up a previously inaccessibly suite of small-scale SNe Ia flame problems.
Shown in Figure~\ref{fig:amr}, for example, is a portion of a calculation of
a Rayleigh-Taylor unstable flame.  The boxes represent the different levels of the adaptive grid hierarchy.  Simulations like this \cite{SNrt,SNrt3d} have
captured the transition to distributed burning in
detail and have shown that the turbulence obeys isotropic Kolmogorov
scaling.

\subsection{The transition to distributed burning}

\begin{figure}[t]
\centering
\begin{minipage}[b]{3.95in}
\includegraphics[width=3.85in]{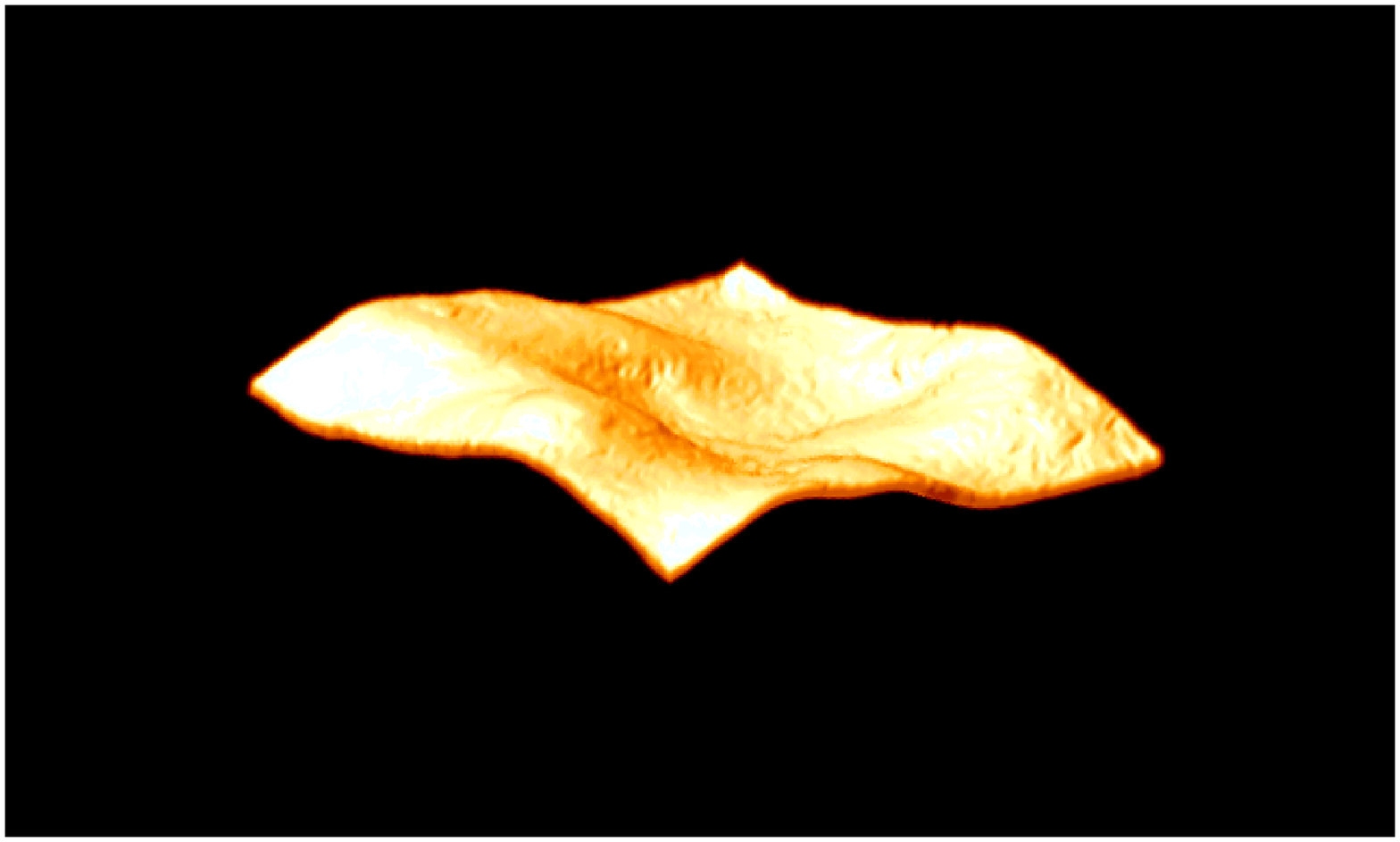} \\[2mm]
\includegraphics[width=3.85in]{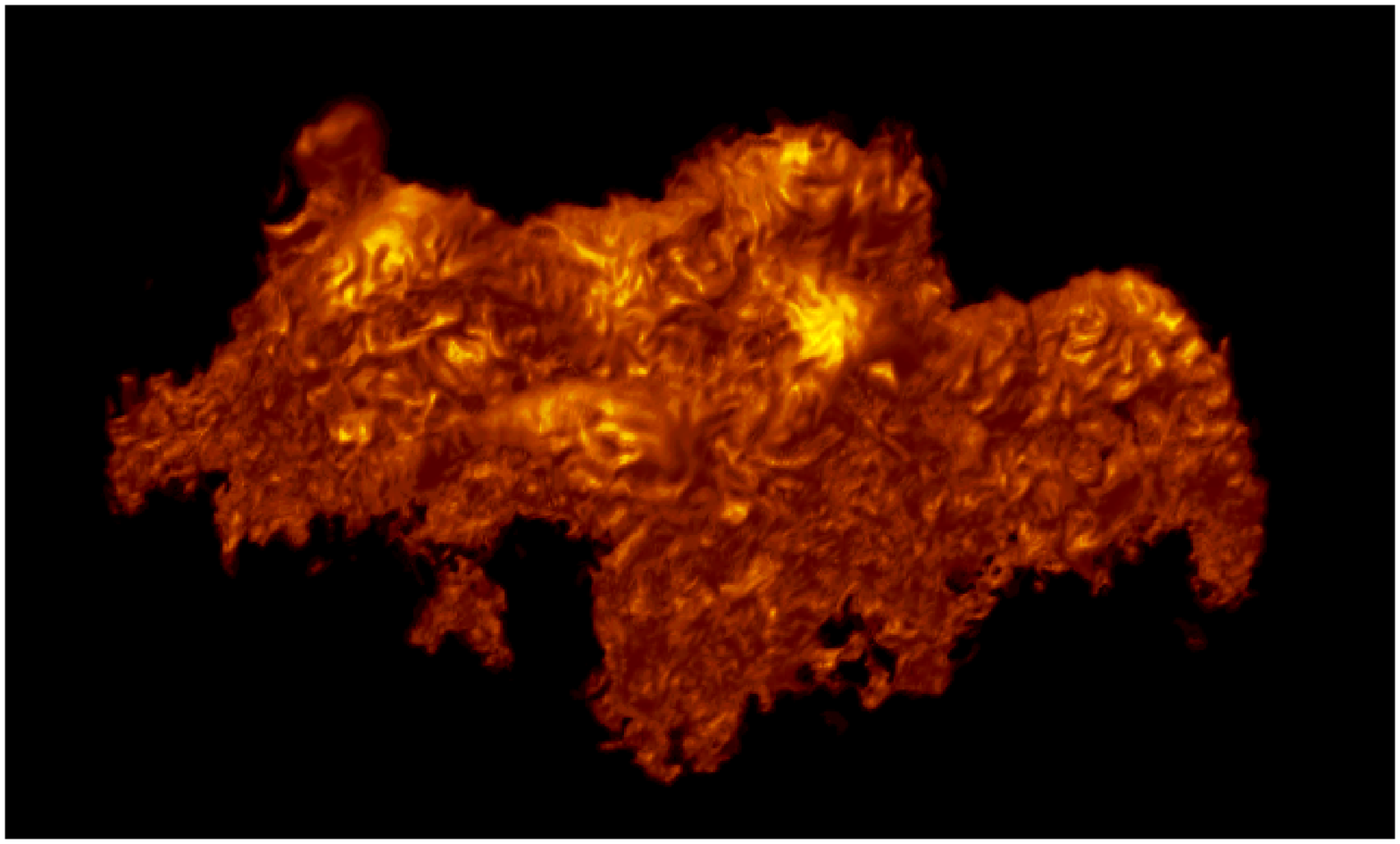}\hspace{0.05in}
\end{minipage}
\begin{minipage}[b]{2.0in}
\caption{\label{fig:turb} Three-dimensional turbulent flames at
$3\times 10^7~\gcc$ (top) and $1.5\times 10^7~\gcc$ (bottom).  In each
case, the domain width is 50 flame thicknesses and the thermal
structure of the flame is resolved.  The turbulent intensity of the
inflowing fuel corresponds to a Kolmogorov cascade from the large
scale motions in the star.\\}
\end{minipage}
\end{figure}

More recent simulations include those that focus on the transition to
distributed burning through a parameter study of turbulent flame
sheets.  Figure~\ref{fig:turb} shows the response of a flame to
inflowing turbulence fuel at two different densities.  In both cases,
the domain width is 50 flame thicknesses in the lateral directions.
The structure of the flame is resolved with 5 zones in its thermal
thickness ($l_f = \Delta T/\max\{\nabla T\}$).  A turbulent velocity
field is generated in an incompressible code and inflowed into the
domain.  The turbulent inflow properties correspond to a Kolmogrov
cascade from an integral scale of $10^6$~cm with a corresponding
velocity of $10^7~\cms$---these numbers are chosen to match the
expected turbulent intensity in SNe Ia \cite{niemeyerwoosley1997}.
Active control is used to vary the inflow velocity to ensure that the
flame remains statistically stationary in the domain as it accelerates.

By varying the density of the flame we can sample different regimes of combustion, from the
flamelet to distributed burning.  At the higher density ($3\times
10^7~\gcc$), with these turbulence parameters, the flame is right
at the point to transition to the distributed regime.  
The flame surface appears relatively smooth, with some large-scale
wrinkling.  The flame can simply burn through the smaller-scale
turbulent eddies.  At the lower density ($1.5\times 10^7~\gcc$), the
flame is wrinkled on a much larger range of scales.  The goal with
this study is to create a model for how large the mixed region of fuel
and ash can grow, and from that, determine whether it is possible for
a transition to detonation. These are the first ever three-dimensional
simulations of distributed burning in SNe Ia.



\section{Full-star low Mach number algorithms}

Addressing the ignition of Type Ia supernovae requires a hydrodynamics
algorithm that can follow the $M \sim 0.01$ convection for many
turnover times on the scale of the full star.  Previous work by
research groups studying convection in SNe Ia includes a
two-dimensional implicit approach \cite{hoflichstein:2002} and use of
the anelastic approximation in three-dimensions
\cite{kuhlen-ignition:2005}.  By contrast, we extend the low Mach
number model described above for small-scale flows to include
compressibility effects due to the stratification of the star.  The
ultimate goal is the development of a new simulation code, {MAESTRO},
that is able to carry the evolution from the convective phase, through
ignition, into the explosion phase.

Following the same procedure as on the small scale, but now allowing
the thermodynamic pressure to vary with time and in the radial direction, i.e.,
$p_0 = p_0(r,t),$ where $r$ is the radial direction and $t$ is time, 
we obtain a constraint on the
velocity field of the form \cite{ABRZ:I,stratified2}:
\begin{equation}
\vec{\nabla} \cdot \vec{U} + \frac{1}{\Gamma_1 p_0} \vec{U}\cdot \vec{\nabla} p_0 = S - \frac{1}{\Gamma_1 p_0} \frac{\partial p_0}{\partial t} \enskip .
\end{equation}
Here, $\Gamma_1 = d (\log p)/ d (\log \rho) |_s$ with $s$ the entropy
of the fluid,  and $S$ is as in (\ref{eq:sn_constraint}).  The
$\vec{U}\cdot \vec{\nabla} p_0$ term represents the compressibility due to
the stratification.
Defining
\begin{equation}
\beta_0(r,t) = \beta(0,t) \exp \left ( \int_0^r \frac{1}{(\Gamma_1 p)_0} \frac{\partial p_0}{\partial r^\prime} dr^\prime \right ) \enskip ,
\end{equation}
(see \cite{ABRZ:I}) we can rewrite the constraint as
\begin{equation}
\vec{\nabla} \cdot ( \beta_0 \vec{U}) = \beta_0 (S - \frac{1}{\Gamma_1 p_0} \frac{\partial p_0}{\partial t})  \enskip .
\end{equation}
The time evolution of $p_0$ is calculated as a hydrostatic adjustment driven
by the the average heating in each radial layer; the details are given in \cite{stratified2}.

\begin{figure}[t]
\centering
\includegraphics[width=6.25in]{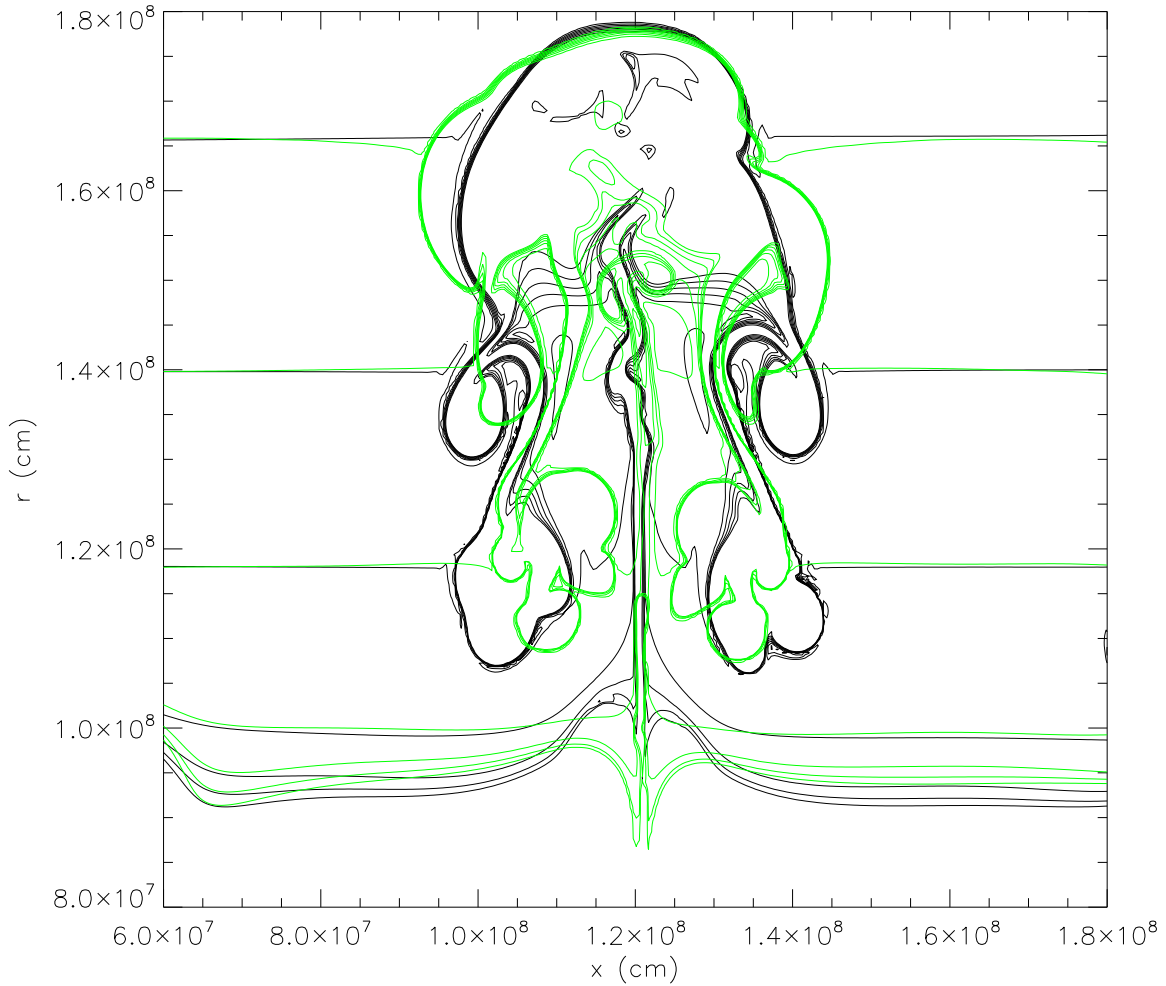} \\[5mm]
\caption{\label{fig:lmn} Comparison of a fully compressible algorithm
(black contours) and the stratified low Mach number algorithm (green
contours) \cite{stratified2}.  Here we focus on a single two-dimensional plume driven
by a localized heating source at $(1.2\times 10^8, 8.5\times
10^8)$~cm.  We see good agreement for the height and width of the
rising plume between the two algorithms.  }
\end{figure}

In the absence of any heating, the background
stratification remains fixed, and the constraint becomes:
\begin{equation}
\vec{\nabla} \cdot ( \beta_0 \vec{U}) = 0 \enskip .
\end{equation}
If we consider an ideal gas, then $\Gamma_1 = \gamma$ (the ratio of
specific heats), and if we take the background to be isentropically
stratified, then this constraint can be written as
\begin{equation}
\vec{\nabla} \cdot (p_0^{1/\gamma} \vec{U}) = \vec{\nabla} \cdot (\rho_0 \vec{U}) = 0 \enskip ,
\end{equation}
which is the constraint used in the anelastic approximation.
Comparisons between the low Mach number approach and the anelastic approximation
\cite{ABRZ:I} show excellent agreement in the regime where the
anelastic approximation is valid.

We note that although the low Mach number formulation reduces to the
anelastic equation set in the case of small-scale heating where the variations in temperature
and density are small, the low Mach number formulation has much
more general applicability.  The anelastic approximation assumes 
a fixed background state and small variations in temperature and density
from that background state.  The stratified low Mach number formulation allows
the background state to vary in time in response to large-scale heating
\cite{almgren:2000,stratified2}, and allows large variations in temperature and 
density from the background values.   

The low Mach number equation set is solved with a second-order
accurate, approximate projection method as in the case of the small-scale
low Mach number approach.  Figure~\ref{fig:lmn} shows
results of a comparison between a fully compressible code and the
stratified low Mach number algorithm for a test case that included
both large-scale and localized heat sources.  
Detailed comparisons between results using the low Mach number approach
and those using fully compressible solvers, both with and
without external heat sources, show excellent agreement between methods. 
As expected, the compressible solver takes many more timesteps than the 
low Mach number method for flows such as this that start from a quiescent state.
The low Mach number approach has been demonstrated to be 
both accurate and efficient for long-time evolution of
astrophysical phenomena \cite{ABRZ:I,stratified2}.

Future development of MAESTRO will focus on formulating the algorithm 
to handle non-grid aligned gravity (i.e.~spherical stars), and improving
the robustness of the algorithm at the edge of the star where the
density drops off suddenly in the radial direction.  In addition
to the target problem of ignition in SNe~Ia, this algorithm can be
applied to classical novae and Type I X-ray burst simulations (see also~\cite{lin2006}).

\section{Summary}

The multiscale nature of Type Ia supernovae makes them challenging 
to simulate numerically.  As for many multiscale problems, advances in
our understanding of the physical phenomena require not just advances
in computer hardware, but new algorithms that respect the physical and
mathematical nature of the phenomena.
This paper outlines several new algorithms for studying SNe Ia
based on the low Mach number approach, and several of the successful 
computations which have resulted.  Future work includes applying these 
algorithms in the hope of answering questions about the nature of ignition in SNe Ia
and whether transition to detonation is possible.

\ack

This work was supported by DOE grant No.\ DE-FC02-01ER41176 to the
Supernova Science Center/UCSC and Applied Mathematics Program of the
DOE Office of Mathematics, Information, and Computational Sciences
under contract No.\ DE-AC03-76SF00098.  The simulations presented were
run on Columbia machine at NASA Ames and the jacquard machine at the
National Energy Research Scientific Computing Center, which is supported by the Office of Science of the U.S. Department of Energy under Contract No. DE-AC02-05CH11231.

\section*{References}

\end{document}